\documentclass[useAMS,usenatbib]{mn2e}
\usepackage{amsmath,amssymb}
\usepackage{natbib}
\usepackage{graphicx}
\usepackage{times}
\def\ifnote{\iffalse}

\title[Lorentz Factor Constraint]{Lorentz Factor Constraint from the very early external shock of the gamma-ray burst ejecta}
\author[Y. C. Zou \& T. Piran]{Yuan-Chuan Zou$^{1,2}$, 
and Tsvi Piran$^{1}$ \thanks{Email: zouyc@hust.edu.cn (YCZ) and
tsvi@phys.huji.ac.il (TP)}
\\
$^{1}${The Racah Institute of Physics, Hebrew University, Jerusalem 91904, Israel}\\
$^{2}${School of Physics,
Huazhong University of Science and Technology,  Wuhan 430074, China}
}

\begin{document}
\date{\today}
\maketitle
\label{firstpage}

\begin{abstract}
While it is generally agreed that the emitting regions in Gamma-Ray Bursts (GRBs) move ultra relativistically  towards the observer, 
different estimates of the initial Lorentz factors, $\Gamma_0$, lead to different, at times conflicting estimates. We show here that
the quiet periods
in which the signals goes down below the  instrumental thresholds,
put strong upper limits on the values of $\Gamma_0$. 
According to the standard internal-external shocks model an external
shock should develop during the prompt stage. This external
shock  radiates in the hard X-rays to soft gamma-rays bands and this emission 
should be seen as a smooth background signal. 
The observed deep minima indicate that this contribution is negligible. This limits, in turn,   $\Gamma_0$.
We obtain  upper limits on $\Gamma_0$ for several bursts with typical values around hundreds. 
We compare these values with those obtained by  the other methods, which 
typically yield lower limits.  The results are marginally consistent leaving  only a narrow range of allowed values for $\Gamma_0$. 
\end{abstract}


\section{Introduction}\label{sec:intr}

The combination of fast variability and a non-thermal spectrum lead to the well known ``compactness problem" and indicates  that the emitting region of GRBs move relativistically towards the observers  \citep{r75} (or that they are extremely nearby).  With the realization, in the early nineties,  that GRBs are cosmological and 
the development of the fireball model, relativistic motion became an essential ingredient of GRB modeling. Relativistic motion was confirmed, first by observations of scintillations at the radio afterglow of GRB 970508  \citep{Goodman97, Waxman98}, later by
afterglow modeling  \citep{spn98, vanp00, kp02} and finally in direct observations of superluminal motion in the afterglow 
of GRB 030329  \citep{tf04,onp04}.  

These observations were all done during the afterglow phase in which the ejected material has already been slowed down by its interaction with the surrounding material and the typical Lorentz factor was of order 5 or less. However, the compactness problem indicates that the initial Lorentz factor, 
$\Gamma_0$,  with which the  jet was ejected from the inner engine and with which it was moving during the prompt GRB phase, was much larger. 

In spite of this progress $\Gamma_0$, which is crucial to understand the underline physics both of the inner engine and of the emission process is still unknown.   Several methods have been proposed to estimate it. The methods vary in complexity and in robustness and depend on different details of the ``internal-external" shocks model and/or on the development of an forward-reverse shocks system during the early phase of the afterglow:
\begin{itemize} 
\item
The simplest, most direct and robust method uses the compactness problem: The  optical depth for high energy photons ($\sim$ GeV) to escape from the emitting region without being annihilated  by softer $\gamma$-rays (sub-MeV) should be less than unity. This leads to a lower limit on the initial Lorentz factor  \citep{feh93, p95, wl95, bh97, ls01}.  It can be used to set an exact value for the Lorentz factor if an upper energy cutoff is observed. However, this was not seen so far. Using this method,  \citet{ls01} obtained  lower limits of $\Gamma_0$  (typically of order of several hundred) for several bursts.

\item  Within the external shocks scenario the peak of the light curve corresponds to the decelerating time of  the ejecta  \citep{mr97, sp99, sp99b, k00}. Observations of this peak provide, therefore, a way to  estimate $\Gamma_0$.  \citet{r09} constrained $\Gamma_o$ using this method  for several  bursts whose prompt optical emission was observed by ROTSE-III.
 \citet{zf06} argued that the  deceleration time should be prior to the shallow decay phase, and used this to obtain lower limits   $\sim 100$ for several bursts.
 
\item  Comparison of the early X-ray and optical emission (that arise  from a  reverse-forward shock system) enables us to estimate the Lorentz factor by fitting parameters of the emitting regions  \citep{sp99,zkm03}. As an example, using this method  \citet{sp99} constrained $\Gamma_0$  of GRB 990123 to be $\sim 200$.   \citet{mv07} and  \citet{jf07} estimated using  the early optical afterglow, $\Gamma_0$,  to be $\sim 400$ for both GRB 060418 and GRB 060607A. Recently,  \citet{xfw09} used the reverse-forward shock model to determine the initial Lorentz factor for several well observed bursts and obtained typical values of 300.

\item  Thermal emission escapes from the fireball's  photospheric  when  it  becomes optically thin.  The observations of a  thermal component would allows us,  therefore, to infer the Lorentz factor at this time  \citep{nps05, pr07}.  Assuming that such a component was indeed observed and using this method  \citet{pr07} estimated $\Gamma_0$  to be hundreds . 

\end{itemize} 

In this work we propose yet another method to estimate $\Gamma_0$.  We work within the ``internal-external" shocks scenario according to which  internal shocks that arise due to  collisions of the ejected shells within the relativistic outflow produce the observed prompt gamma-rays, while the interaction of the merged shells with  the surrounding medium produces the afterglow  \citep[see][for a review]{p05}. However, an external shock begins to develop even during the prompt phase. It is caused by the outermost shell that sweeps up the external medium. This external shock radiates and produces an underlying  smooth component. Usually, this extra component will be sub-dominant  when the internal shocks emission is strong. 
However, as   this emission is not observed even during troughs of the light curves  it must be weaker than the 
detection threshold in cases when  a deep minimum in the light curve is observed.  For example, {\it Swift} BAT has a sensitivity limit of	 $\sim 10^{-8} {\rm ergs\,cm^{-2}\,s^{-1}}$  \citep{gc04} (corresponding to $f_{\nu,{\rm lim}} \sim 4 \times 10^{-28} {\rm ergs\,cm^{-2}\,s^{-1}\,Hz^{-1}}$). 
We show here that the strength the early forward shock emission depends sensitively on the Lorentz factor at this stage (which is roughly the initial one). Using this we set upper limits on the initial Lorentz factor for  variant bursts. 

We derive the radiation flux density from the forward shock, and the resulting constraint on $\Gamma_0$ in section  \ref{sec:model}. We examine other methods for constraining $\Gamma_0$  in section \ref{sec:other}. We calculate the constraints for several selected individual bursts using three methods in section  \ref{sec:cases} and discuss the implications of the results in section \ref{sec:con}.

\section{Model}\label{sec:model}

Our model is based on a basic ingredient of the ``internal-external shocks model".  Within this model, while internal shocks are going on producing the 
prompt gamma-rays the outermost shell that is at the front of the ejecta begins interacting with the surrounding matter and  an 
a reverse-forward shocks system develops    \citep{sp95,zkm03,np04}. The reverse shock propagates back into the front of the ejecta and the external shock propagates into the surrounding matter. We consider here the emission from this very early reverse-forward shocks system. The contribution of the forward shock should appear in soft gamma-rays as a smooth and continuous emission with an increasing signal. However, in many cases the observed signal decreases to very low values, which can be even below the detection limit of the observing instrument. We use this to constrain the initial Lorentz factor. Depending on the environment, we consider two cases: interstellar medium (ISM) and a wind.

\subsection{An ISM}

The reverse shock can be relativistic (RRS) or Newtonian (NRS) depending on a density ratio between the ejecta and the surrounding matter. The relevant case depends on the parameter  \citep{sp95,np04}:
\begin{equation}
\xi \equiv \left(\frac{l}{\Delta_0}\right)^{1/2} \Gamma_0^{-4/3} 
  \simeq 34 \mbox{n}_{0}^ {- {{1}\over{6}} }\,\Delta_{0,10}^ {- {{1
 }\over{2}} }\,\eta_{2}^ {- {{4}\over{3}} }\,E_{53}^{{{1}\over{6}}},
 \label{eq:xi}
\end{equation}
where $l \equiv (3E/4 \pi n m_p c^2)^{1/3}$ is the Sedov length, $\Gamma_0$ is the initial Lorentz factor of the merged shell, which is also the Lorentz factor of the prompt gamma-ray emitting region, $\Delta_0$ is the initial width of the shell, which is related to the duration of the pulse by $\delta t \sim \Delta_0/(2c) \sim 0.1 s$  \citep{wd03}, $E$ is the isotropic kinetic energy, $n$ is the matter number density, and $m_p$ is the proton rest mass. The conventional notation $Q=Q_x \times 10^x$ is used throughout this paper. $\xi < 1$ leads to RRS, while $\xi > 1$ corresponds to NRS.

Typical values of the parameters have been used in eq. (\ref{eq:xi}). Unless $\eta$ or $\Delta$ are very large, $\xi \gg 1$, so we only consider the NRS case. As the Newtonian reverse shock radiates near the optical band, the main contribution to the X-ray to soft gamma-ray emission arises from the forward shock, which we examine now.

The deceleration radius for the outermost shell is
\begin{equation}
R_d = \left( \frac{3 E_0}{4\pi n \Gamma_0^2 m_p c^2} \right)^{1 \over 3} =
1.2\times 10^{17} E_{53}^{1\over 3} \Gamma_{0,2}^{- {2\over 3}} n_0^{- {1\over 3}}
{\mbox{cm}}
\end{equation}
Consequently, the deceleration time is
$
 t_{\oplus,d} = \frac{R_c}{2\eta^2 c} \simeq 1.9\times 10^{2} (1+z) E_{53}^{\frac{1}{3}} \Gamma_{0,2}^{-\frac{8}{3}} n_0^{-\frac{1}{3}} {\mbox{s}}
$, which is much longer than the time of the first minimum in which we are interested. So we can consider the scaling-laws for the very early external shock before it has been decelerated. During this phase the shell coasts with a constant Lorentz factor $\Gamma_0$, collecting the medium to radiate X-rays and the shock accelerated electrons are cooled mainly via synchrotron emission.

Following \citet{spn98}  \citep[see also in e.g.][]{np04,zwd05}, 
 we calculate the dynamics and radiation from the forward shock.
The internal energy density of the forward shocked material is
$e = 4 \eta^2 m_p c^2 n \simeq 60\,n_{0}\,\Gamma_{0,2}^2 \,{\mbox{erg}\,\mbox{cm}^{-3}}$, 
the magnetic field in the comoving frame is
$B = \sqrt{8\pi \epsilon_B e} \simeq 12 \,n_{0}^{{{1}\over{2}}}\,\epsilon_{B,-1}^{{{1
 }\over{2}}}\,\Gamma_{0,2} {\, \rm Gauss}$,
where $\epsilon_B$ is the equipartition factor for the magnetic energy density.
The peak spectral power is
$P_{\nu,\max} = (1+z) \sigma_T m_e c^2 \Gamma_0 B/ (3 q_e)
 \simeq 4.7 \times 10^{-19}\,(1+z)\,n_{0}^{{{1}\over{2}}}\,
 \epsilon_{B,-1}^{{{1}\over{2}}}\,\Gamma_{0,2}^2 {\, \rm erg \, Hz^{-1} \, s^{-1} }$,
 where $\sigma_T$ is the Thomson cross section, $q_e$ is the electron charge.
The peak observed flux density is then
$f_{\nu,\max} = {N_e P_{\nu,\max} }/{(4\pi D^2)}
 \simeq 3.4 \times 10^{-31}\,D_{28}^ {- 2 }\,(1+z) ^ {- 2 }\,
 n_{0}^{{{3}\over{2}}}\,t_{\oplus}^3\,\epsilon_{B,-1}^{{{1}\over{2}}}
 \,\Gamma_{0,2}^8  {\, \rm erg \, cm^{-2} \, Hz^{-1} \, s^{-1}}$,
 where $N_e$ is the total number of emitting electrons, and $D$ is the luminosity distance. 
The synchrotron cooling Lorentz factor is
$\gamma_c = 6 \pi m_e c/(\sigma_T B^2 t_{co})
\simeq 2.6 \times 10^4 \,\mbox{n}_{0}^ {- 1 }\,\Gamma_{0,2}^ {- 3 }\,(1+z)\,t_{\oplus}
 ^ {- 1 }\,(1+Y)^ {- 1 }\,\varepsilon_{B,-1}^ {- 1 }
$
(where $t_{co}$ is the comoving time scale and $Y$ is the Compton parameter for synchrotron self-Compton scattering).
This corresponds to the cooling frequency
$\nu_c = (1+z)^{-1} \frac{\Gamma_0 \gamma_c^2 q_e B}{2 \pi m_e c}
 \simeq 2.3 \times 10^{18}\,(1+z)\,n_{0}^ {- {{3}\over{2}} }
 \,t_{\oplus}^ {- 2 }\,\epsilon_{B,-1}^ {- {{3}\over{2}} }\,
 \Gamma_{0,2}^ {- 4 } {\, \rm Hz}$.
The typical Lorentz factor of the electrons is
$\gamma_m = \frac{p-2}{p-1} \frac{\epsilon_e e}{n_2 m_e c^2}
 \simeq 2.0 \times 10^5\,\zeta_{{{1}\over{3}}}\,\Gamma_{0,2}\,\varepsilon_{e,-{{1}\over{2}}}
$, 
and the typical synchrotron frequency is
$\nu_{m} = (1+z)^{-1} \frac{\Gamma_0 \gamma_m^2 q_e B}{2 \pi m_e c}
 \simeq 1.3 \times 10^{18}\,(1+z)^ {- 1 }\,n_{0}^{{{1}\over{
 2}}}\,\epsilon_{B,-1}^{{{1}\over{2}}}\,\epsilon_{e,-0.5}^2\,
 \Gamma_{0,2}^4\,\zeta_{1/3}^2 {\, \rm Hz}$,
where $\zeta = 3 \frac{p-2}{p-1}$, and $p$ is the index of power law distributed electrons.
The Synchrotron-self absorption frequency is
$\nu_{a} \sim 2.8 \times 10^{8}\,(1+z)^ {- {{13}\over{5}} }\,n_{0}^{{{9}\over{5}}
 }\,t_{\oplus}^{{{8}\over{5}}}\,\epsilon_{B,-1}^{{{6}\over{5}}}\,
 \Gamma_{0,2}^{{{28}\over{5}}} {\, \rm Hz}$ for $\nu_a < \nu_c < \nu_m $
and
$\nu_{a} \sim 4.6 \times 10^{8}\,(1+z)^ {- {{8}\over{5}} }\,n_{0}
 ^{{{4}\over{5}}}\,t_{\oplus}^{{{3}\over{5}}}\,\epsilon_{B,-1}^{{{1
 }\over{5}}}\,\epsilon_{e,-0.5}^ {- 1 }\,\Gamma_{0,2}^{{{8}\over{5}}}
 {\, \rm Hz}$ for
$ \nu_a < \nu_m < \nu_c$, which are always well below $\nu_m$ and $\nu_c$.

Collecting the above expressions,  the observed flux density $f_\nu$ is:
\begin{eqnarray}
f_\nu &=& 2 \times 10^{-33}\,\Gamma_{0,2}^9\,D_{28}^ {- 2 }\,(1+z)^ {- {{9
 }\over{4}} }\,n_{0}^{{{9}\over{8}}}\,t_{\oplus}^2
\epsilon_{B,-1}^{
 {{1}\over{8}}}\,\epsilon_{e,-\frac{1}{2}}^{{{3}\over{2}}}\,
 \nu_{20}^ {- {{5}\over{4}} } \nonumber \\
&& (1+Y)^{-1} {\, \rm erg\, cm^{-2} Hz^{-1} s^{-1}}
\label{eq:fnu}
\end{eqnarray}
 for $p=2.5 $\footnote{Note that for simplicity, the above relations are given for $p=2.5$.  For other values of $p$, the indices will change accordingly but not too much. For example, if $p=2.2$, the scaling law is $f_\nu \propto \Gamma_0^{42/5}$ rather than $ \propto \Gamma_0^9$. As the flux density in the gamma-ray band is $f_{\nu,\max} (\nu_c/\nu_m)^{-(p-1)/2}(\nu/\nu_c)^{-p/2}$ or  $f_{\nu,\max} (\nu_c/\nu_m)^{-(p-1)/2}(\nu/\nu_c)^{-p/2}$, which is lower with higher value of $p$. Therefore, the choice of $p=2.5$ is conservative when evaluating the minimal flux density, 	as the usual value of $p$ is around 2.2 \citep{a01}. Furthermore, the value of $p$ in the range of [2,3] doesn't change the emission much. See also eq. (\ref{eq:etalim-p}) for a general $p$.} and  $\nu > (\nu_a, \nu_c, \nu_m)$. 
One can see that $f_\nu$  depends very sensitively on the initial Lorentz factor $\Gamma_0$. This is partially because of the number of emitting electrons $N_e$, for $N_e = 4\pi/3 R^3 n \propto \Gamma_0^6 t_\oplus^3 n$, behaves like $\Gamma_0^6$.

\begin{figure}
 \includegraphics[width=0.5\textwidth]{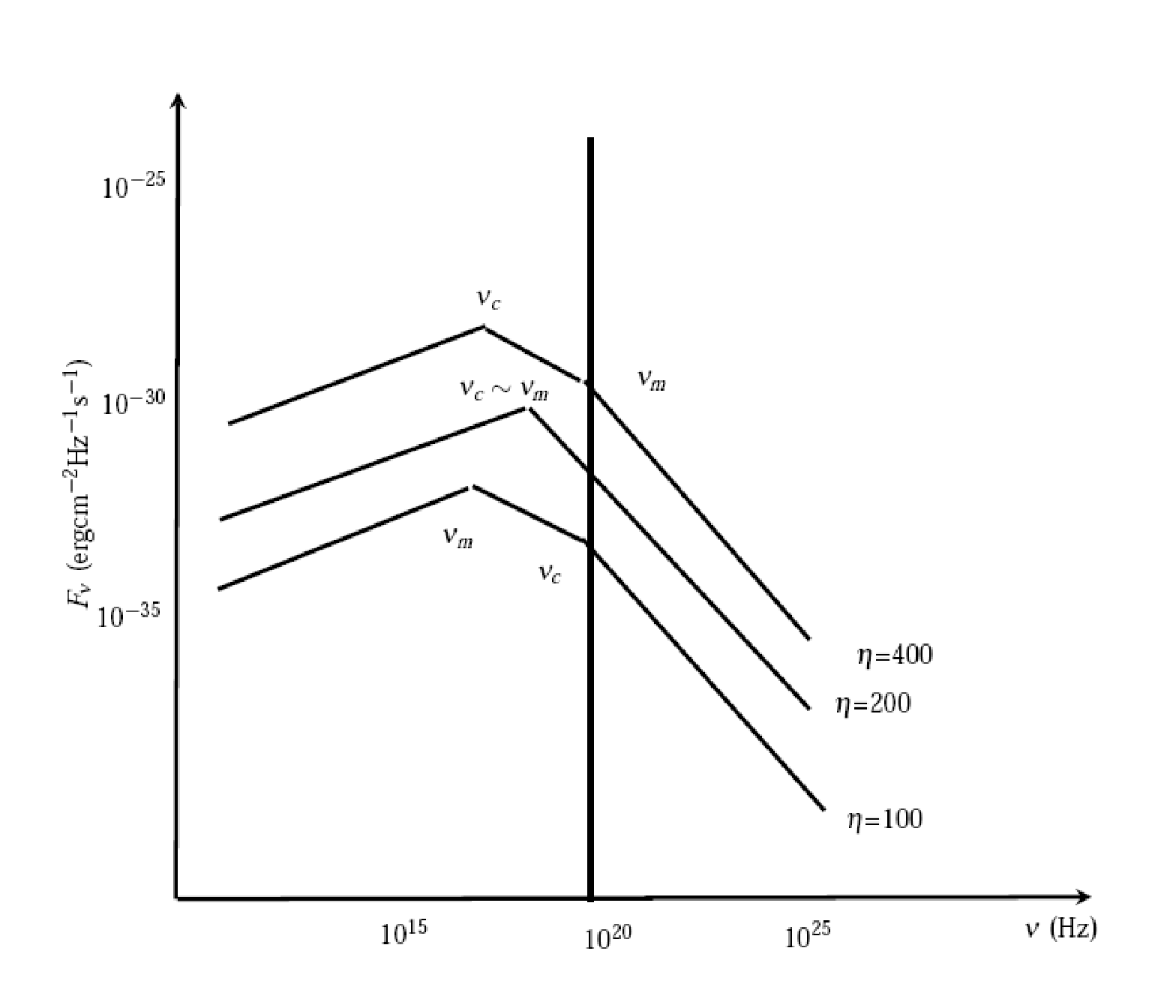}
 \caption{The sketched synchrotron spectra of the forward shock emission before the deceleration. The lines indicate for $\Gamma_0$ = 400, 200 and 100 from top to down respectively.}
 \label{fig:fnu}
\end{figure}

For different parameters, especially for different values of $\Gamma_0$, $\nu$ may be either greater or smaller than $\nu_m$ or $\nu_c$. Figure \ref{fig:fnu} shows rough spectra and the relations between the different frequencies for 3 different values of $\Gamma_0$. Even though the observed frequency is not always larger than $\nu_c$ and $\nu_m$, it is never significantly smaller. Therefore, equation (\ref{eq:fnu}) is roughly acceptable for the whole range of parameters of interest. For simplicity, we use only this equation in the following discussion.

If the observed flux density immediately after the first GRB pulse is $f_{\nu,C}$, The emission of the early external shock should not exceed this value, requiring $f_{\nu} \leq f_{\nu,C}$. In other cases, there is no signal immediately after the first pulse, when $f_{\nu}<f_{\nu,lim}$, where $f_{\nu,lim}$ is the limiting flux density of the observing instrument. We use $f_{\nu,lim}$ to represent both quantities in the following.
Using $f_{\nu,lim} = 10^{-28} {\rm erg\,cm^{-2}\,Hz^{-1}\,s^{-1}}$,
 this inequality gives a strong constraint on $\Gamma_0$:
\begin{eqnarray}
\Gamma_0 < 340 (1+z)^{\frac{1}{4}} f_{\nu,lim,-28}^{\frac{1}{9}} D_{28}^{\frac{2}{9}} n_0^{-\frac{1}{8}} 
   \epsilon_{e,-\frac{1}{2}}^{-\frac{1}{6}} \epsilon_{B,-1}^{-\frac{1}{72}} \nu_{20}^{\frac{5}{36}} t_{\oplus}^{-\frac{2}{9}} (1+Y)^{\frac{1}{9}},
\label{eq:etalim}
\end{eqnarray}
Note that the limit practically depends rather weakly on all the other parameters.  For completeness, we also present the expression for a genrael $p$: 
\begin{eqnarray}
\Gamma_0 &< & 340 \times 2.4^{-\frac{p-2.5}{9(p+2)}} \left[f_{\nu,lim,-28} \, (1+z)^{\frac{p+2}{2}}  D_{28}^{2} \right.
\nonumber \\ 
 && n_0^{-\frac{p+2}{4}} 
   \epsilon_{e,-\frac{1}{2}}^{-(p-1)} \epsilon_{B,-1}^{-\frac{p-2}{4}} \nu_{20}^{\frac{p}{2}} \, t_{\oplus}^{-2} \nonumber \\ 
&& \left.[{3(p-2)}/{(p-1)}]^{-(p-1)}(1+Y)\right]^{\frac{1}{2p+4}}.
\label{eq:etalim-p}
\end{eqnarray}
As the value of p ranges between 2 and 3 generally, and as the overall expression  depends on $\frac{1}{2p+4}$, $\Gamma_0$ is weakly dependent on the other parameters for a general $p$.

\subsection{A wind environment}
For a wind environment, the distinction between RRS and NRS depends on  \citep{wd03}
\begin{equation}
 \xi \equiv \left(\frac{l}{\Delta_0}\right)^{1/2}\Gamma_0^{-2} \simeq 23 E_{53}^{1/2} A_\star^{-1/2} \Delta_{0,10}^{-1/2} \Gamma_{0,2}^{-2}
\end{equation}
where the Sedov length in a wind case is $l  \equiv E_0/( 4 \pi A m_p c^2)$, $A$ is the wind parameter satisfying $n(r)=A r^{-2}$, and $A=3\times 10^{35} A_\star {\rm \, cm^{-1}}$. For typical parameters the NRS case is also most likely.
The deceleration radius is 
\begin{equation}
R_d = \frac{ E_0}{4\pi A \Gamma_0^2 m_p c^2}  =
1.8\times 10^{16} E_{53} \Gamma_{0,2}^{- 2} A_{\star,-1}^{- 1}
{\mbox{cm}},
\end{equation}
and the corresponding deceleration time is $t_{\oplus,d}=30 (1+z) E_{53} \Gamma_{0,2}^{- 4} A_{\star,-1}^{- 1}$sec, 
which is also longer than the duration of a typical first pulse. 
The observed flux density of the forward shock before deceleration is:
\begin{eqnarray}
f_\nu &\sim& 2 \times 10^{-27} \,\Gamma_{0,2}^{{9}\over{2}}\,D_{28}^ {- 2 } \,A_{\star,-1}^{{{9}\over{8}}}\,t_{\oplus}^{-{{1}\over{4}}}
\epsilon_{B,-1}^{ {{1}\over{8}}}\,\epsilon_{e,-\frac{1}{2}}^{{{3}\over{2}}}\, \nu_{20}^ {- {{5}\over{4}} }  
\nonumber \\ 
&&
(1+Y)^{-1}{\rm \, erg\, cm^{-2} Hz^{-1} s^{-1}} 
\end{eqnarray}
in the case $\nu>(\nu_m,\nu_c,\nu_a)$ (see  \citet{zwd05} for other cases).
As the wind density is much higher than the ISM density, the flux density is much higher than the one expected for an ISM (given by eq. (\ref{eq:fnu})). For the typical parameter $A_\star=0.1$, the number density at a given radius $R=2 \Gamma_0^2 c t_\oplus \sim 6\times 10^{14} \Gamma_{0,2}^2 t_\oplus$ cm is about $10^6$ larger than the ISM, and consequently, the flux density is about  $10^6$ stronger, as it is proportional to $n^{9/8}$.
For $p=2.5$, the initial Lorentz factor can be expressed as
\begin{equation}
\Gamma_0 < 50 f_{\nu,lim,-28}^{\frac{2}{9}} D_{28}^{\frac{4}{9}} A_{\star,-1}^{-\frac{1}{4}}   \epsilon_{e,-\frac{1}{2}}^{-\frac{1}{3}} \epsilon_{B,-1}^{-\frac{1}{36}} \nu_{20}^{\frac{5}{18}} t_{\oplus}^{\frac{1}{18}} (1+Y)^{\frac{2}{9}}.
\label{eq:etalim-wind}
\end{equation}
The limit is even stronger than the one obtained in ISM case. This  is essentially because of the denser medium for the typical wind parameter. However, the density  profile is uncertain at  small values of $r$ (that are relevant at this stage) and it is not clear if this limit is valid.  In any case, this is a strong argument against very dense environment near GRBs.

\section{Other constraints}\label{sec:other}
As mentioned earlier several other methods to constrain the initial Lorentz factor have been suggested. We compare the constraints obtained using  the different methods to show the consistency of the overall model. As most other methods give lower limits, combined with ours, we obtain a overall  stronger constraint.

\subsection{Compactness and the optical depth for pair production}\label{sec:LS}
Observations of high energy photons from a GRB imply that the emitting region is optically thin for pair production of these high energy photons with  the lower energy  sub-MeV gamma-rays. This leads to a limit \citep{ls01} (denoted limit A)
\begin{equation}
 \Gamma_0 > \hat{\tau}^{1/(2\beta+2)}(E_{\max}/m_e c^2)^{(\beta-1)/(2\beta+2)}(1+z)^{{\beta-1}/{\beta+1}},
 \label{LS01}
\end{equation}
where $\beta$ is the photon spectral index ($f E^{-\beta} {\rm {cm}^{-2} {s}^{-1} {erg}^{-1}}$), $E_{\max}$ is the maximal energy of the observed high energy photons, and $\hat{\tau}$ is defined as 
\begin{eqnarray}
 \hat{\tau} &\equiv& \frac{(11/180)\sigma_T D^2(m_e c^2)^{-\beta+1}f}{c^2 \delta T (\beta-1)} \nonumber \\
 &\simeq & 4.3\times 10^{10} \frac{f_1}{\beta-1}\,  D_{28}^2 \, 0.511^{-\beta+1} \delta T_{-1}^{-1} ,
\end{eqnarray}
where $D$ is the luminosity distance, $\delta T$ is the variability time scale of the prompt emission, and $f_1 \equiv f ({\rm MeV})^{-\beta+1} {\rm s^{-1} cm^{-2} }$ is the number of photons per second per square centimeter per MeV at energy of 1 MeV.

When high energy observations are not available, we can still obtain a bound by assuming that the observed higher energy spectrum at around an MeV can extend up to the energy where photons can just annihilate with themselves in the same spectral shape.
This leads to another constraint  \citep{ls01} (denoted limit B):
\begin{equation}
 \Gamma_0 > \hat{\tau}^{1/(\beta+3)}(1+z)^{{\beta-1}/{\beta+3}}.
 \label{LS01B}
\end{equation}
Typically, limit A yields a tighter constraint than limit B.

\subsection{The deceleration time}\label{sec:dec}
The deceleration time of the external forward shock, $t_{\gamma,2}$ corresponds to the  peak in the  afterglow light curve.
It depends weakly on the overall energy and the external density and most sensitively on the 
initial Lorentz factor.  At the deceleration time, the Lorentz factor is   a  half of the initial Lorentz factor. This yields  \citep{sp99b}:
\begin{equation}
 \Gamma_0 \sim 190 E_{k,52}^{1/8} n_0^{-1/8} t_{\gamma,2}^{-3/8}(1+z)^{3/8}.
\end{equation}
where $t_{\gamma}$ is the deceleration time, that corresponds to the peak in the afterglow light curve.

By employing an efficiency parameter $\bar{\eta} \equiv E_{\rm \gamma,iso}/E_{\rm k,iso}$, where $E_{\rm k,iso}$ and $E_{\rm \gamma,iso}$ are isotropic equivalent kinetic energy and the prompt emitted photon energy respectively, we get 
\begin{equation}
 \Gamma_0 \sim 220 E_{\rm \gamma,iso,52}^{1/8}{\bar{\eta}_{-0.5}}^{-1/8} n_0^{-1/8} t_{\gamma,2}^{-3/8}(1+z)^{3/8},
 \label{eq:SP99}
\end{equation}

For a wind environment, the initial Lorentz factor is  \citep{pk00}
\begin{equation}
 \Gamma_0 \sim 110 E_{\rm \gamma,iso,52}^{1/4}{\bar{\eta}_{-0.5}}^{-1/4} A_{\star,-1}^{-1/4} t_{\gamma,2}^{-1/4}(1+z)^{1/4},
 \label{eq:SP99b}
\end{equation}
where $A_\star=3\times 10^{35} {\rm cm^{-2}}$ is the wind parameter.


\subsection{A photospheric thermal component}
A thermal component,  should come from the fireball's photosphere (where the optical depth is unity). Detection of such a component would enable us to  determine the initial Lorentz factor using   \citep{nps05, pr07}:
\begin{equation}
 \Gamma_0 \simeq \left[ (1+z)^2 D \frac{F_{obs}\sigma_T}{2 m_p c^3 \mathcal{R}}\right]^{1/4},
\end{equation}
where $\mathcal{R}=(F_{BB}/\sigma T_{obs}^4)^{1/2}$, and $F_{BB}$ is the thermal component emission.

\subsection{The reverse-forward shocks system}
During the early afterglow, the predicted reverse-forward shock may give a direct clue to determine the initial Lorentz factor, as the Lorentz factor of the unshocked ejecta (that is involved with the reverse shock) is   roughly equal to the initial Lorentz factor. By fitting the early afterglow to the predictions of the reverse-forward shock model, one obtains different parameters within the shocked region including  the Lorentz factor. This is the final Lorentz factor after the merging during internal shocks, which may be regarded as an average value of the initially ejected sub-shells.

\ \\

We have listed several methods. However, not all are practical. Specifically, 
 it is difficult to dig out a thermal component within the total emission as needed for  the thermal component method \citep[see however][]{r05}. Similarly, the identification of 
a reverse-forward shocks system is  limited and available only for a few bursts.  In the following we will use only the methods in sections \ref{sec:LS} (LS01A or LS01B) and \ref{sec:dec} (SP99).

\section{Case study and a comparison of the different methods}\label{sec:cases}

We examine some well observed bursts to see whether the different constrains on the initial Lorentz factor   are consistent with each other.
The new constraint described here is based on the observation of the first pulse and the following minimum.  For subsequent pulses  the forward shock 
is complicated. Therefore, the constraint is only suitable for those bursts which exhibit a short first pulse followed by a deep trough.  In the following we consider only such bursts.  To use the three methods, we need the bursts have the properties: (i) A GeV signals (for LS01A). (ii)  A clear  first pulse and  a subsequent minimum (for this work, ZP09), with not long duration, otherwise, the external shock may have decelerated during the first pulse. (iii) A peak in the optical afterglow light curve (for SP99). However, it is very rare  satisfying all the three requirements. Then we reduce it to any burst who has two of the above preperties as the sample to examine the $\Gamma_0$. We use typical parameters: $f_{\nu,{\rm lim}} \sim 4 \times 10^{-28} {\rm ergs\,cm^{-2}\,s^{-1}\,Hz^{-1}}$ as the detection limit if it is a {\it Swift} burst; $\bar\eta \simeq 0.3$,  $n=1 {\rm cm^{-3}}$, $p=2.5$, $\epsilon_e \simeq 0.3$, and $\epsilon_B=0.1$ if they are not determined.

\ifnote
With early optical data, but not good to use.

main origin for data searching: 
Rykoff 2009, ROTSE data; 
Kann 2007(arXiv:0712.2186), {\it Swift} optical afterglows;
Kann 2005(astroph:0509466), catalog before {\it Swift}.

\subsection*{GRB 980613(no)}
Afterglow too late, 18 hr. 

\subsection*{GRB 980703(no)}
Afterglow too late, 1day. 

\subsection*{GRB 990510(no)}
Afterglow too late, 0.1day. 

\subsection*{GRB 990712(no)}
Afterglow too late, 0.1day. 

\subsection*{GRB 991208(no)}
Afterglow too late, 100,000s. 

\subsection*{GRB 000301C(no)}
Afterglow too late, 0.2days. 

\subsection*{GRB 010222(no)}
Afterglow too late, 0.2days. 

\subsection*{GRB 011211(no)}
Afterglow too late, 0.5days. 

\subsection*{GRB 020405(no)}
Afterglow too late, 1 days. 

\subsection*{GRB 020813(no)}
Afterglow too late, 0.1 days. 

\subsection*{GRB 021004 (no)}
It was taken reverse-forward shock. 

\subsection*{GRB 021211 (no)}
GRB 021211 was located at 
The first optical data was taken $130$secafter the burst, and the light curve decayed since that  \citep{li03}, which implies the decelerating time is shorter than 130sec. 
But it may also be reverse-forward shock. 

\bibitem[Li et al.(2003)]{li03}Li W., Filippenko A. V., Chornock R., \& Jha S., 2003, ApJ, 586, L9

\subsection*{GRB 030115 (not good)}
Good pulse shape (but only one pulse for the burst). 
Opt is too late (2 hours). 

\subsection*{GRB 030226 - bad}
GRB 030226 was located at z=1.98. Both uniform medium or wind-type medium can fit the observed muti-band light curves, with parameters ...
But the prompt light curve is not suitable for ZP09.

GRB 011121 is not good neither with the same reason(2004ApJ...616.1078F).

\subsection*{GRB 030328 (no)}
First time of optical observation is OK, 0.06days. 

But the prompt light curve is poor. The first pulse is too long. 

\subsection*{GRB 030329 (not good)}
Good shaped prompt pulses. 
Opt is too late (~1000s). 

\subsection*{GRB 030418 (no)}
No redshift.

\subsection*{GRB 031203(no)}
Afterglow too late, 10 days. 

\subsection*{GRB 041006 (maybe no)}
HETE burst.

first optical data at 0.04 days. But may be reverse-forward shock.

\subsection*{GRB 050315 (no)}
No good shaped first pulse. 

\subsection*{GRB 050315 (no)}
Good shaped first pulse. 
Opt is too late, 5000s. 

\subsection*{GRB 050319 (no)}
This burst was located at $z=3.24$, with $T_{90} = 150$sec, fluence in the 15 - 350 keV band $f \simeq 1.6 \times 10^{-6} {\rm ergs cm^{-2}}$.

the duration of the pulse were appeared 100s before: http://gcn.gsfc.nasa.gov/notices\_s/111622/BA/ 

So  not easy to carry out the ZP09 method.

\subsection*{GRB 050408}
Good light curve shape. 
Optical peak less than 0.03 day. 

\subsection*{GRB 050410 (no)}
Pulse too long. 
Maybe dark burst.

\subsection*{GRB 050412 (no)}
Not good valley after the first pulse. 
Maybe dark burst.

\subsection*{GRB 050416a}
Good pulse. 
Not good optical observations. We can't see what is the decelerating time, nor even the lower limit of decelerating time. 

\subsection*{GRB 050502A (no)}
GRB 050502A was located at $z=3.793$ ($D_L\sim 1.1 \times 10^{29}$cm)  \citep{pe05}. 
I can't find the prompt light curve.
The first optical observation was at $44$sec, which implies to be the peak time from the light curve  \citep{yost06}. The $\gamma$-ray isotropic equivalent energy is . Giving the typical parameter, we get the upper limit to be $650$ by method ZP09, and the value of initial Lorentz factor to be $??$ by method SP99.

\bibitem[Prochaska et al.(2005)]{pe05}Prochaska J. X., Ellison S., Foley R. J., Bloom J. S., \& Chen H.-W. 2005, GCN Circular, 3332
\bibitem[Yost et al.(2006)]{yost06}Yost S. A., et al., 2006, ApJ, 636, 959

\subsection*{GRB 050525a (no)}
Optical afterglow may be modeled by reverse-forward shock (Shao \& Dai 2006), which is different from pure afterglow.

\subsection*{GRB 050603 (no)}
Good light curve. 
Opt data too late. 5000s. 

\subsection*{GRB 050730 (no)}
Bad light curve. 

\subsection*{GRB 050802 (no)}
No good first pulse. 

\subsection*{GRB 050820a (no)}
no redshift. Gamma light curve not clear. 

Good optical data. 
But obviously energy injection.

\subsection*{GRB 050824 (no)}
Bad light curves. 

\subsection*{GRB 050908 (no)}
Good single pulse (not so good). 
But no published opt obs. 

\subsection*{GRB 05128 (no)}
Only one long pulse (maybe external shock produced gamma ray burst). 
Opt observations began at 0.1 days. 

\subsection*{GRB 051109a}
This burst was located at $z=2.346$ ($D_L\sim 5.9 \times 10^{28}$ cm)  \citep{quimby05}. The duration of the first pulse is $\sim 9$sec(http://gcn.gsfc.nasa.gov/notices\_s/163136/BA/). For typical parameters, the upper limit is $\sim 490$ by method ZP09. The lower limit by method SP99 is $\sim 600$.

\bibitem[Quimby(2005)]{quimby05}Quimby R., Fox D., Hoeflich P., Roman B., & Wheeler J. C. 2005, GCN Circ., 4221

\subsection*{GRB 051111 (no)}
The duration of the pulse is too long. (Maybe itself is the external shock, which should be confirmed with x-rays.)

\subsection*{GRB 060111b (no)}
redshift not determined. 

\subsection*{GRB 060117 (no)}
Good light curve of gamma-rays. 
However, the optical dots depends on models. Maybe a reverse-forward shock. 
No redshift determined

\subsection*{GRB 060124 (no)}
Pulse is OK. 
But the opt is too late (0.1 day). 

\subsection*{GRB 060206 (no)}
bad early optical observations. The late flare can't tell anything about the decelerating.

\subsection*{GRB 060210 (no)}
A pulse at time ~-100s, don't know how to treat it.
The peak at ~1000s doesn't look like a pure fireball decelerating.

\subsection*{GRB 060418 (no)}
The first big pulse is too long, then the forward shock may be decelerated already. 

\subsection*{GRB 060502A (no)}
Good shaped single pulse, but it is too long(30s). 

\subsection*{GRB 060512, 060707, 060714, 060729, 061121, 070110, 070411, 070419A, 070612A, 070810A (no)}
Pulse is OK, or good, or ... 
No published Opt data. 

\subsection*{GRB 060526 (no)}
Only one point at early time, which can't tell it is decelerating or coasting.
Good prompt light curves. 

\subsection*{GRB 060605 (no)}
The first pulse is 20s, which is too long.

\subsection*{GRB 060729 (not good)}
Two peaks in the optical light curve. Not easy to determine.

\subsection*{GRB 060904b (not good)}
This burst was located at $z=0.7$ ($D_L \sim 1.3 \times 10^{28}$cm). The first pulse lasted about 30secand reached the detectability of {\it Swift} BAT  \citep{klotz08}. So the maximal Initial Lorentz factor from ZP09 is $\Gamma_0 \le 230 \,n_0^{-1/8} \epsilon_{e,-1/2}^{-1/6} \epsilon_{B,-1}^{-1/72}$. However, if the external shock had decelerated before the 30sec, this constraint does not hold. From the optical peak, the initial Lorentz factor is derived as $\sim 310$  \citep{klotz08}.

\bibitem[Klotz et al(2008)]{klotz08}Klotz A., et al., 2008, A\&A, 483, 847

\subsection*{GRB 060908 (not good)}
There is a clear valey at about 10 s. 
No published optical data, though the GCNs hints the decelerating time should be less than 0.01day. 

\subsection*{GRB 060927 (not good)}
Good first pulse.
Opt, there is a decay at the end of the prompt stage, and a peak then. Not easy to see either have decelerated before the decay, or decelerated at the peak time. 

\subsection*{GRB 061126 (not good)}
No well shaped first pulse. 
z=1.1588.
Optical may be from reverse-forward shock. 

\subsection*{GRB 070125 (no)}
Good gamma ray data. 
But no early afterglow. 

\subsection*{GRB 070420 (no)}
Optical peak at 200sec  \citep{klotz08}. 
No good shaped single pulse.

\subsection*{GRB 070611 (no)}
Poor shape.

\subsection*{GRB 080129 (no)}
Can't distinguish the afterglow from optical flares.

\subsection*{GRB 081126 (no)}
Good gamma-ray shape, good optical peaks, but z not determined.
\fi

\subsection*{GRB 990123}
GRB 990123 was located at $z=1.60$ ($D_L=3.7\times 10^{28}$ cm) with an isotropic equivalent energy in $\gamma$-rays $E_{\gamma,\rm iso} \sim 3.4 \times 10^{54}$ ergs  \citep{k99}. It has been extensively discussed. 
 \citet{ls01} obtained a lower limit of $\Gamma_0 \approx  150$ using   LS01A and 180 using LS01B (assuming there is a single power law in the spectrum of the high energy band).

The afterglow light curve can be fitted both by a uniform medium and a wind-type medium, with parameters  $n=0.004 {\rm cm^{-3}}$, $\epsilon_e=0.075$, $\epsilon_B=4\times 10^{-4}$, for uniform medium and  $A_\star=0.06$, $\epsilon_e=0.08$, $\epsilon_B=7\times 10^{-5}$,
for wind-type medium respectively, while the underlying physical parameters are somewhat uncertain  \citep{panaitescu05}. 
The prompt BATSE light curve shows a valley at $\sim 12$sec \citep{g99}.
BATSE sensitivity in 50-300 keV,  is $0.2 {\rm \, cm^{-2} s^{-1}}$ (http://glast.gsfc.nasa.gov/science/instruments/table1-2.html), corresponding to a flux density of  $ f_{\nu,lim} \sim 1.25 \times 10^{-27} {\rm erg\,cm^{-2} s^{-1} Hz^{-1} }$.
Using those parameters and equations (\ref{eq:etalim}) and (\ref{eq:etalim-wind}), we obtain  an  upper limit of $\Gamma_0 <  1200$ for a uniform medium and 410 for wind-type medium by our method.

From the optical light curve  \citep{g99}, the decelerating time should be less then 0.1 day ($\sim 8640$ s). (Though there were also optical observations before, they are generally considered as reverse-forward shock signals  \citep{sp99}.) Using equations (\ref{eq:SP99}) and (\ref{eq:SP99b}), and assuming the efficiency $\bar{\eta}=0.3$, we obtain  lower limit of  100  for uniform medium and 130  for wind-type medium respectively by method SP99.

These constraints are consistent with those obtained  by light curve fitting,  e.g.   $\Gamma_0 = 270$ by  \citet{k00} and $\Gamma_0 = 1200$ by  \citet{wdl00}.

\subsection*{GRB 021004}
GRB 021004 was observed by HETE-II, and located at redshift $z=2.32$ ($D_L\sim 5.8\times 10^{28}$ cm) \citep{fox03}. 
Lacking  high energy data, we cannot method LS01A.
To use method LS01B, we need the spectrum of the prompt emission. However, the spectrum fitting was a single power-law with photon index  $\beta \sim 1.64$  \citep{lamb02}. An extended single power-law with $\beta < 2$ indicates the most energy is hidden in the higher energy band and there must be a spectral break and an unknown high energy spectral index. Therefore,  we cannot carry out the estimation by method LS01B neither.

The duration of the first pulse is $\sim 10$sec (http://space.mit.edu/HETE/Bursts/GRB021004/). HETE's Detection thresholds for the French Gamma Telescope (FREGATE) is $\sim 3\times 10^{-8} {\rm erg \, cm^{-2} \, s^{-1}}$ (http://space.mit.edu/HETE/fregate.html), which corresponding to $f_{\nu,lim} \sim 1.2 \times 10^{-27} {\rm erg \, cm^{-2} \, Hz^{-1}\, s^{-1}}$ at 100 keV. The afterglow emission was fitted well by wind-type environment, with parameters $A_\star =0.6, E_{k,52} = 10, \epsilon_e = \epsilon_B=0.1$ \citep{lc03}. Using equation (\ref{eq:etalim-wind}), we obtain the upper limit of $\Gamma_0 < 210$.

From the optical light curve, which was first observed 537sec after the trigger of the burst  \citep{fox03}, the decelerating time of the external shock should be shorter than this time.  Using equation (\ref{eq:SP99b}), we obtain (by method  SP99) $\Gamma_0 > 80$.

\subsection*{GRB 040924}
GRB 040924 was a short burst with duration $T_{90} = 2.39 \pm 0.24$ s. It   was located at $z=0.858$ ($D_L \sim 1.7 \times 10^{28}$ cm)  \citep{wv05}. 
 The peak energy of the time-integrated spectrum obtained by Konus-wind was $E_p = 67 \pm 6$ keV. However, the spectral indices were not available  \citep{ga04}. Therefore, method LS01 can not be used.

The $\gamma$-ray isotropic equivalent energy was $ \sim 1.5 \times 10^{52}$ erg  \citep{fz05}. The first optical observations was taken at $\sim 1000$sec \citep{wv05}, which yields an  upper limit of decelerating time.
As the parameters are not firmly determined  \citep{fz05}, we choose the typical parameters (with HETE-II limit: $f_{\nu,lim} \sim 1.2 \times 10^{-27} {\rm erg \, cm^{-2} \, Hz^{-1}\, s^{-1}}$). We find  $\Gamma_0 < 490$ by method ZP09, and $\Gamma_0 > 120$ using SP99.

\subsection*{GRB 050401}
GRB 050401 was located at $z=2.9$ ($D_L \sim 7.6 \times 10^{28}$ cm). The duration of the first pulse was $\sim 6$sec \citep{dep05}. 
From Konus-Wind observation  \citep{ga05}, the 2nd peak of the prompt pulses had $E_p = 119 \pm 26$ keV, $\alpha = 0.83$, $\beta=2.37$ and peak flux $2.45 \pm 0.12 \times 10^{-6} {\rm erg cm^{-2} s^{-1}}$ (in the 20 keV - 2 MeV energy range). Without a direct high energy ($\sim$GeV) detection, we use method LS01B to get a lower limit. Using  $f_1 \sim 0.27 \, {\rm cm^{-2} s^{-1} MeV^{-1}}$ and $\hat \tau \sim 1.7 \times 10^{10}$, we get $\Gamma_0 >110$. 
Using the BAT limit and other typical parameters, and  ZP09 we obtain $\Gamma_0 < 590$. The lower limit   \citet{r09} using method SP99 is $\Gamma_0 > 900$.

\subsection*{GRB 050801}
GRB 050801 was located at $z=1.56$ ($D_L\sim 3.6\times 10^{28}$ cm).
There was no observed emission at time $\sim 8$sec \citep{dep07}. 
The peak flux was $1.7 \pm 0.1 {\rm ph\, cm^{-2} s^{-1}}$ in 15-350 keV at 1sec, and the time averaged  spectral index was $\beta = 2.0 \pm 0.2$  \citep{s05} (we use this as the index at the peak time). Taking the duration of the first pulse to be 2sec, we get $f_1 \sim 0.03 \, {\rm cm^{-2} s^{-1} MeV^{-1}}$, $\hat \tau \sim 	1.5\times 10^9$, and $\Gamma_0 > 80$ using   LS01B.
Giving the typical parameter values of $n$, $\epsilon_e$, and $\epsilon_B$, we get $\Gamma_0 < 420$ using  ZP09. The lower limit using  SP99 is $\Gamma_0 > 500$  \citep{r09}.

\subsection*{GRB 050922c}
GRB 050922c had  $T90 \sim 4.5$s  \citep{sakamoto08}. It was located at $z=2.198$ \citep{jakobsson05} ($D_L\sim 5.5\times 10^{28}$ cm), with fluence $f= 3.1 \times 10^{-6} {\rm erg \, cm^{-2}}$ in 30-400 keV  \citep{crew05}, which  corresponds to $E_{\rm iso} \sim 3.7 \times 10^{52}$ erg. 
As the spectral power-law index was $\beta = 1.55 \pm 0.07$ ($<2$)  \citep{ga05b}, we cannot use method LS01.
Taking  T90 as the duration of the first pulse and  the typical parameters, we get $\Gamma_0 < 550$ using  ZP09. A lower limit  using method SP99 is  $\Gamma_0 > 350$ \citep{r09} .

\subsection*{GRB 060607a}
GRB 060607a was located at $z = 3.082$ ($D_L \sim 8.2 \times 10^{28}$ cm)  \citep{nr09}. 
With no high energy observation and a  BAT photon spectral index $\beta =1.45 \pm 0.07$ we cannot use method LS01.
The first peak ended at $\sim 15$sec \citep{zh08}. The isotropic equivalent energy was $E_{\rm iso} \sim 1.1 \times 10^{53}$ erg. A clear optical peak was observed at $\sim 180$sec \citep{nr09}. With typical parameters, the upper limit using  ZP09 is $\Gamma_0 <  490$. The inferred initial Lorentz factor using SP99 is $\sim 410$.

\subsection*{GRB 060614}
GRB 060614 was located at redshift $z=0.125$ ($D_L=1.8\times 10^{27}$cm) with an isotropic equivalent energy $E_{\gamma,\rm iso} \sim 2.5 \times 10^{51}$ ergs  \citep{m07}. 
There was an exponential cut-off at $\sim 300$ keV for the intense pulse  \citep{ga06}, which means there are practically no  photons at higher energy.  So we can not use method LS01. 
The duration of the first pulse was about 1sec  \citep{gc06}. From afterglow modelling, the physical parameters are $E_k\sim 6\times  10^{50}$ ergs, $\epsilon_e \sim 0.12$, $\epsilon_B \sim 0.0002$, and $n=0.04 {\rm cm^{-3}}$  \citep{xu09}. Taking the instrument limit of {\it Swift} BAT,
we find (using  ZP09)  $\Gamma_0 < 530$. 
The peak time of the afterglow was $\sim 3\times 10^4$ sec  \citep{m07}. Taking this as the decelerating time, and $\bar \eta = 0.3$, we find (using  SP99) $\Gamma_0 \sim 35$. This pretty low value may imply that the peak of the optical afterglow for GRB 060614 did not occur at  the deceleration time but it arised due to energy injection as suggested by  \citet{xu09}.

\subsection*{GRB 061007}
This burst was located at $z=1.26$ ($D_L \sim 2.7 \times 10^{28}$ cm). 
The duration of the first pulse was $\sim 3$sec \citep{o09}. 
The peak time was at $\sim 39.5$sec, with $E_p = 498 ^{+54} _{-48} $ keV, $\alpha=0.53 ^{+0.08}_{-0.09}$, $\beta = 2.61 ^{+0.49} _{-0.25}$, and the peak flux $1.95 ^{+0.31} _{-0.24} {\rm erg \, cm^{-2} s^{-1}}$  \citep{ga06b}. Taking the duration of the peak pulse to be 0.3sec(see http://gcn.gsfc.nasa.gov/notices\_s/232683/BA/ for the prompt light curves), we get $f_1 \sim 3.5 \, {\rm cm^{-2} s^{-1} MeV^{-1}}$, $\hat \tau \sim 6.7\times 10^{11}$, and $\Gamma_0 >160$ using   LS01B.

The micro-physics parameters being uncertain\citep{m07,s08}, we choose the normal value $\epsilon_e=0.3, \epsilon_B=0.1$ and $n=1 {\rm cm^{-3}}$, which don't affect the result much. Using the {\it Swift} BAT limit and equation (\ref{eq:etalim}), we find $\Gamma_0 \le 480 \,n_0^{-1/8} \epsilon_{e,-1/2}^{-1/6} \epsilon_{B,-1}^{-1/72}$ with ZP09.
The optical afterglow peaked at $\sim 39$sec, and the isotropic equivalent gamma-ray energy was $\sim 1.4 \times 10^{54}$ erg  \citep{r09}. We obtain, using equation (\ref{eq:SP99}), $\Gamma_0  \approx 640 \, \bar{\eta}_{-0.5}^{-1/8}n_0^{-1/8}$, which is consistent with  \citet{r09}.

The later two constraints are inconsistent, and there is no much space to tune the parameters and reach consistency. However, one have to recall that method SP99 is somewhat crude and it is not clear that the inconsistency of less than a factor of 2 is significant. Moreover, a possible explanation for this contradiction could be that the Lorentz factor of the first ejected shell is less than 460, while the followed other shells move faster than the first one. After they merged, produce a single shell moving with a higher Lorentz factor. This may be an evidence that the outermost shell is accelerated during the prompt phase by other shells.

\subsection*{GRB 080319B}
GRB 080319B, the naked eye burst,   was located at redshift $z=0.937$  \citep{Vreeswijk08}. Its duration
$T_{90}$ was $\sim 57$sec. The peak flux was $F_p \sim 2.26 \pm 0.21 \times 10^{-5} {\rm erg \,cm^{-2}s^{-1}}$ and the peak of the $\nu F_{\nu}$
spectrum was $E_p \simeq 675{\pm 22}$keV (i.e., $\nu_p \sim 1.6\times 10^{20}$Hz,  consequently $F_{\nu,p} \sim 1.4 \times 10^{-25} {\rm erg\, cm^{-2} Hz^{-1} s^{-1}}$).  The photon indexes below and above  $E_p$ are $0.855^{+0.013}_{-0.014}$ and
$3.59_{-0.32}^{+0.62}$ respectively  \citep{r08}. With luminosity distance $D\sim 1.9 \times 10^{28}$cm, GRB 080319B had  a peak
luminosity $L_p \sim 9.67\times 10^{52} {\rm erg}\,{s}^{-1}$ and an
isotropic equivalent energy $E_{\rm \gamma, iso} \simeq 1.32 \times 10^{54}$
erg  \citep{ga08}.
The peak photon fluence is roughly $\sim 20 {\rm cm^{-2} s^{-2}}$  \citep{r08}. Taking the duration of pulse $\sim 2$sec, 
we get $f_1\sim 2.3 \, {\rm cm^{-2} s^{-1} MeV^{-1}}$, $\hat \tau \sim 3.9 \times 10^{10}$ and $\Gamma_{i,\min} \sim 50$ using method LS01B.
The duration of the first pulse was about 3sec, and the valley after the pulse was about 1 mJy  \citep{r08}. With   $A_\star=0.01, \epsilon_e=0.2, \epsilon_B \sim 6 \times 10^{-7}$  \citep{r08}, we find,   using ZP09, $\Gamma_0 < 580$.
From the optical light curves  \citep{p09}, one can estimate the deceleration time of the afterglow to be less than 100sec. Setting the efficiency parameter $\bar{\eta}=0.3$, and the wind parameter $A_\star=0.01$  \citep{r08}, we find, using  SP99, $\Gamma_0 > 810$.

\subsection*{GRB 080916C}
This burst was located at $z \sim 4.35$ ($D_L \sim 1.2 \times 10^{29}$ cm). To carry out method LS01, we focus on the 2nd pulse, which took place between  3.6 - 7.7sec.  Two photons with energy $>1GeV$ were detected during this pulse.  For the soft gamma-ray band, the peak energy was $\sim 1170$ keV, the photon spectral index in the higher band was $\sim 2.21$, and the peak flux density was $3.5 \times 10^{-2} {\rm \, cm^{-2} s^{-1} {keV}^{-1}}$ \citep{a09}, and then the flux at 1 MeV was $f_1 \sim 48  {\rm \, cm^{-2} s^{-1} MeV^{-1}}$.  We find $\hat\tau \sim 1.34 \times 10^{13}$ and $\Gamma_0 > 870$ using LS01A. This  is consistent with the result  of   \citet{a09}, while $\Gamma_0 > 490$ by method LS01B.

Consider the first pulse of this burst, which was during the period $0.004 - 3.58$sec \citep{a09}. The external forward shock should not exceed the observed average gamma-ray flux $\sim 6.9 {\rm \,cm^{-2} s^{-1}}$  \citep{a09} in energy band $50 - 300$ keV. As the photon index was $\sim 0.58$, the flux density at 300 keV was $\sim 1.6 \times 10^{-26} {\rm {ergs\, cm^{-2}\, Hz^{-1} \, s^{-1}}}$, which can be taken as a conservative $f_{\nu,lim}$. With typical parameters and using eq. (\ref{eq:etalim}), we get,  using ZP09, $\Gamma_1 <  1130$.
The optical afterglow showed that the deceleration time was less than $6\times 10^4$s  \citep{greiner09}. With $E_{\rm \gamma,iso}\sim 8.8 \times 10^{54}$ ergs  \citep{a09}, we get the initial Lorentz factor $\Gamma_0 > 90$ using SP99.

\ifnote
With High energy data, but not good to use.

\subsection*{GRB 080825C}
All  the LAT events detected during the GBM emission have energies below 1  GeV.

(bad data description, GCN 8183)

\subsection*{GRB 081024B}
This burst's redshift is not available.
The emission from this point source was seen up to 3 GeV  \citep{Omodei09}

\bibitem[Omodei et al.(2009)]{Omodei09}Omodei N., et al., 2009, GCN circular, 8407

\subsection*{GRB 090217}
This burst was found in a blind search in the LAT data,
and also independently by follow up studies around the GBM location.
The emission continues for up to 20 seconds after the GBM trigger,
 but commences several seconds after the GBM trigger.

(LAT data not clear: GCN 8903)

\subsection*{GRB 090323}
Emission was observed in the LAT up to a few
GeV. The high-energy emission commences several seconds after the GBM
trigger time, and we see marginal evidence in the LAT that it continues
for up to a couple of kilo-seconds  \citep{Ohno09}.

(The GeV emission is unclear: GCN 9021, not good for LS01
The 1st pulse is too long: This time interval consists of two main emission peaks, from T0-2.0sec
to T0+33.8sec and from T0+33.8sec to T0+71.7 s. GCN 9035, not good for ZP09.
All the optical data were taken after 1 day, e.g., GCN 9026. Not good for SP99.)
\bibitem[Ohno et al(2009)]{Ohno09}Ohno M., et al., 2009, GCN circular, 9021
\fi

\subsection*{GRB 090328A}
This burst was located at $z=0.736$  \citep{Cenko09} ($D_L \sim 1.4\times 10^{28}$ cm). 
The spectrum from T0+3.1sec to T0+29.7sec was best fitted by a Band function 
with indices $\alpha=0.93 \pm 0.02$ and $\beta=2.2 \pm 0.1$, and peak energy 
of $E_{\rm peak}=653 \pm 45$ keV.
The fluence in this time interval was $8.09 \pm 0.10 \times 10^{-5} {\rm erg\, cm^{-2}}$ 
in the 8-1000 keV band and $9.5 \pm 1.0 \times 10^{-5} {\rm erg\,cm^{-2}}$ in the 8keV-40MeV 
band. The isotropic equivalent energy in the 8keV-40MeV band was 
$E_{\gamma, \rm iso} = 2.3 \pm 0.2 \times 10^{53}$ ergs. The 1-sec peak photon flux measured 
starting from T0+23.5sec in the 8-1000 keV band was $18.5 \pm 0.5 { \, \rm ph\,s^{-1}\,cm^{-2}}$  \citep{rau09a}, which corresponds to $f_1 \sim 2.7 { \, \rm ph\,s^{-1}\,cm^{-2} \, MeV^{-1}}$, and therefore $\hat\tau \sim 4.2 \times 10^{10}$ (taking the variability time scale  to be 1sec). 
The Fermi Large Area Telescope (LAT) had  detected this GRB with emission  observed up to a few GeV \citep{McEnery09}. However, the arrival time was very uncertain, even up to 900sec \citep{Cutini09}. Here we assume the observed highest photon is 5 GeV, and it was in the same time interval of the prompt soft $\gamma$-rays. We  get  $\Gamma_0 \ge 320$ using LS01A and  $\Gamma_0 \ge 130$ using LS01B.
The first pulse was about 4.2sec \citep{rau09a}. By taking typical parameter, we find $\Gamma_0 < 540$ using ZP09.
As the early full optical light curve is not available, we cannot carry out the SP99 method for this burst.

\subsection*{GRB 090424}
GRB 090424 was located at $z=0.544$  \citep{chornock09} ($D_L = 9.6 \times 10^{27}$ cm). 
The peak energy of this burst was $E_p = 177 \pm 3$ keV, with $\alpha = 0.9 \pm 0.02$ and $\beta = 2.9 \pm 0.1 $  \citep{connaughton09}. The 0.128-sec peak photon flux measured
at 1.4sec in the 8-1000 keV band was $137 \pm 5 {\rm ph\,s^{-1}cm^{-2}}$ (pulse duration $\sim 0.3$ s
)  \citep{connaughton09}. We get $f_1 \sim 1.6 \, {\rm cm^{-2} s^{-1} MeV^{-1}}$, $\hat \tau \sim 4.0 \times 10^{10}$, and $\Gamma_0 \ge 70$ using LS01B. Without observations of  high energy (GeV) photons , method LS01A cannot be used.

The fluence (8-1000 keV) over the entire event was $5.2 \pm 0.1 \times 10^{-5} {\rm erg \, cm^{-2}}$  \citep{connaughton09}, corresponding to the $E_{\gamma,iso} \sim 4 \times 10^{52}$ erg. We take the valley at 6sec (the first major pulse was $\sim 6$sec with a few sub-pulses  \citep{connaughton09}) to constraint the initial Lorentz factor using ZP09. Using typical parameters, we get $\Gamma_0 \le 300$. 

The optical temporal  index varied from $\sim 1.2$ during $\sim 100 - \sim 1000$sec \citep{xin09}. This indicate deceleration time should be less than $\sim 100$sec. Using method SP99, we get the $\Gamma_0 \ge 310$.

\subsection*{GRB 090510}
GRB 090510 was classified as short burst with a duration $0.5$sec  \citep{h09}. It was located at a redshift $z=0.903 \pm 0.003$  \citep{rau09}, corresponding to a luminosity distance $1.8 \times 10^{28}$ cm. The integrated spectrum was well fitted by a Band function with
$\alpha = 0.80 \pm 0.03$, $\beta = 2.6 \pm 0.3$ and $E_{p} = 4.4 \pm 0.4$ MeV, and the 8 keV to 40 MeV fluence was  $\sim 3.0 \times 10^{-5} {\rm\, erg\,cm^{-2}}$  \citep{g09}, corresponding to the total isotropic equivalent energy $6.4 \times 10^{52}$ erg. The peak photon flux was $ 80 \, {\rm cm^{-2}s^{-1}}$, corresponding to $f_{4.4} \sim 24\, {\rm cm^{-2} s^{-1} MeV^{-1}}$ (at 4.4 MeV) and $f_1 \sim 204 \, {\rm cm^{-2} s^{-1} MeV^{-1}}$ (which is the extension from the $\beta$ slope), so $\hat{\tau}\sim 5.2 \times 10^{13}$. There were $> 10$ photons with energy $>$ 1 GeV during the prompt phase  \citep{Omodei09}. Using this photons  and taking  the duration of the sub-pulses is 0.1sec, the minimal Lorentz factor satisfies $\Gamma_0 >  960 $ using  LS01A and  $\Gamma_0 > 340$ with LS01B. The minimal energy to annihilate the high energy photon is $E_{\max, an}=(\Gamma m_e c^2)^2/E_{\max} \sim 26 \Gamma_3^2 E_{\max,10GeV}^{-1}$MeV, which is well above the break point of the soft-gamma spectrum. Therefore, the photon index 2.6 can be safely used.

The first valley occurred at $\sim 0.1$sec. Using equation (\ref{eq:etalim}), we find $\Gamma_0 <  620$ using  ZP09. UVOT found an optical peak at $\sim 600$sec \citep{Kuin09}. Taking  this as the deceleration time of the external shock, and assuming $n=1 {\rm cm^{-3}}$, $\bar{\eta}=0.3$, we obtain  $\Gamma_0 \sim 180$ using SP99.

\begin{table}
\caption{Comparison of the initial Lorentz factor constraint of different methods. The inconsistent cases are in boldface.}
\begin{tabular}{llllll}
 \hline
Burst No.& z & SP99 & LS01  & ZP09 & medium\\
\hline
GRB 990123 & 1.60 & $\ge 100$ & $\ge 180^\ddag$ & $\le 1200$ & uniform \\
                     & $-$ & $\ge 130$& $-$& $\le 410$& wind-type \\
GRB 021004 & 2.32 & $\ge 80 $ & a,b & $\le 210$ & wind-type\\
GRB 040924 & 0.858 & $\ge 120$  & a,c & $\le 490$ & uniform\\
{\bf GRB 050401} & 2.9 & $\ge 900$ & $\ge 110^\ddag$ & $\le 590$ & uniform\\
{\bf GRB 050801} & 1.56 & $\ge 500$  & $\ge 80^\ddag$ & $\le 420$ & uniform\\
GRB 050922C & 2.198 & $\ge 350$  & a,b & $\le 550$ & uniform\\
GRB 060607A & 3.082 & $\sim 410$  & a,b & $\le 490$ & uniform\\
GRB 060614 & 0.125 & $\sim 35$  & a,d & $\le 530$ & uniform\\
{\bf GRB 061007} & 1.26 & $\sim 640$  & $\ge 160^\ddag$ & $\le 480$ & uniform\\
{\bf GRB 080319B} & 0.937 & $\ge 810$  & $\ge 50^\ddag$ & $\le 580$ &wind-type\\
GRB 080916C & 4.35 & $\ge 90$ & $\ge 880^\dag$  & $\le 1130$ & uniform\\
GRB 090328A & 0.736 & e & $\ge 320^\dag$  & $\le 540$ & uniform\\
{GRB 090424} & 0.544 & $\ge 310$  & $\ge 70^\ddag$  & $\le 300$ & uniform\\
{\bf GRB 090510} & 0.903 & $\sim 180$ & $\ge 960 ^\dag$ & $\le 620$ & uniform\\
\hline
\multicolumn{6}{|l|}{a. no high energy ($\sim$GeV) observations; b. photon index is less than 2; }\\
\multicolumn{6}{|l|}{c. no photon indices; d. exponential cutoff; e. no optical data available.}\\
\multicolumn{6}{|l|}{$^\dag$ limit A of LS01; $^\ddag$ limit B of LS01.}\\
\hline
\end{tabular}
\label{tab:LFs}
\end{table}

\ \\

\begin{figure}
 \includegraphics[width=0.5\textwidth]{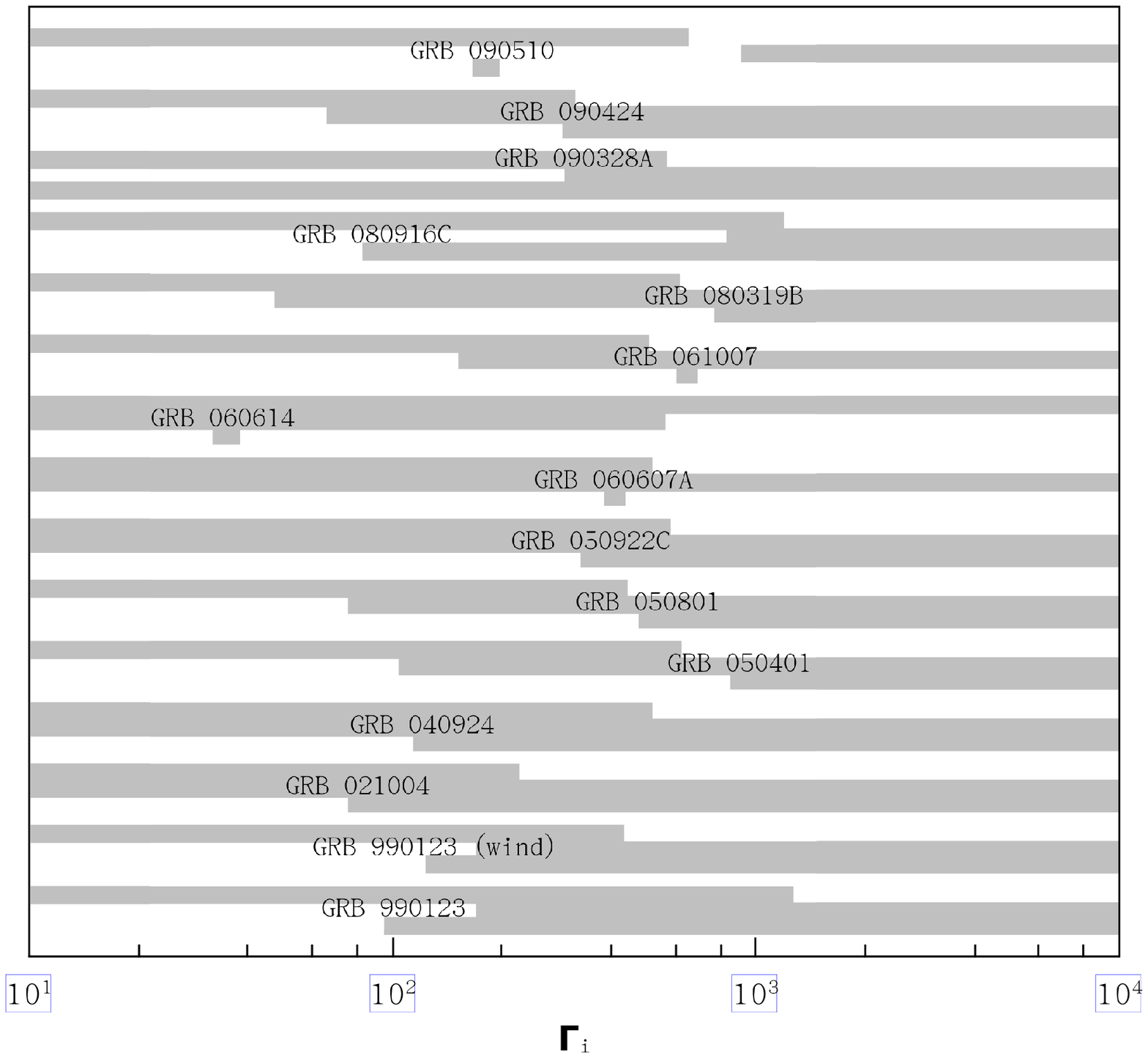}
 \caption{The allowed space (in grey) for $\Gamma_0 $ for different bursts by three different methods. For each burst, from bottom up the methods are SP99, LS01 and ZP09 respectively. For the limits A and limit B in method LS01, only the higher value is used.}
 \label{fig:LFs}
\end{figure}

The limits on  the initial Lorentz factor obtained using  different methods are summarized in Table \ref{tab:LFs} and depicted graphically in fig. \ref{fig:LFs}. 
For a few  bursts (GRBs 050401, 050801, 061007, 080319B, 090424, 090510) the limits  are inconsistent. One may wonder if this inconsistency is problematic. 
 First, we should realize the ``initial Lorentz factor" for the different methods point to different objects. In the method SP99, it is the ``final" Lorentz factor after all the sub-shell merged. In  method LS01 it corresponds to the specific  shell which produces the  GeV photons. In our method, ZP09,  it is the Lorentz factor of  the  first shell. It is possible that different objects have different Lorentz factors even for the same event.

Moreover, the methods that depend on different  assumptions, may not be that accurate. First, all  three constraints assume the relation $R=2\Gamma_0^2 c \delta t$ to obtain the  emission radius. Method  SP99, assume no energy injection, and it depends on  parameter such as the  density $n$ and  the gamma ray efficiency $\bar{\eta}$. When using  LS01, $E_p$ should be less than $E_{\max, an}$, and the high energy spectrum  should obey the Band function. Finally  for ZP09, particle's power-law distribution and early formed external shock are assumed.

\section{Conclusion and discussion}\label{sec:con}

We have considered the very early emission from the external shock, that begins to develope already during the prompt stage and  
emits soft $\gamma$-rays. Such emission was not identified so far. It is possible that we have  seen such early soft $\gamma$-ray emission from the external shock, either as a single smooth light curve like the one observed in GRBs 911022 and 920216 \citep{f94}, or as a smooth pulse following some additional pulses like GRB 050525A \citep{bb06} and 080916A \citep{z09}, or that the signal was hidden among numerous pulses like GRBs 911106 and 911127, 920221 \citep{f94}.   However, one cannot confirm that these signals arose  from external shocks. On the other hand, in many bursts a clear strong minima, reaching the sensitivity limit of the detector that follows the first pulse puts a very strong limits on this emission. 
These  limits constrain  the physical parameters of these GRBs and in particular  the initial Lorentz factor. In cases when a clear minima is seen after the first prompt pulse this  leads to a strong upper limit, typically  of order of a few hundreds on the Lorentz factor. The exact value  depends rather weakly  on the sensitivity of the observing instrument, the distance, the density of the environment, the duration of the first $\gamma$-ray pulse and on the micro-physical parameters.

In view of the insensitivity of the constraint to various parameters, it is
 rather robust, provided that it is applicable and that the external shocks model is relevant at this stage.   For example  the very early external shock results from the interaction with matter that is rather close to the progenitor star and the environment is rather uncertain. 
These considerations rule out a dense
$r^{-2}$ wind profile that extends all the way to
small distance from the progenitor as this
would produce a too strong early external
shock signal.
Additionally, it applies to the Lorentz factor of the outermost shell, which could be  slow relative to subsequent shells that follow. 

A comparison with two other independent methods to estimate the Lorentz factor reveals that the three methods are inconsistent  (by a factor of up to 1.5) in 5 out of 14 cases considered.  This factor of 1.5 may not be significant taking into account the uncertainties in some of the methods. 
In method SP99, the Lorentz factor depends on the total kinetic energy and the environmental density, while the kinetic energy is very uncertain.
In method LS01, the Lorentz factor depends sensitively on the spectral index $\beta$ which is uncertain  especially in the higher energy band. The assumption of a spectral single power-law in the higher energy, used in LS01B,  may also be invalid. 
Moreover, we should notice that the different methods actually address  different ``initial Lorentz factors": the ``final" Lorentz factor after all the sub-shells merged (in SP99), the Lorentz factor of the shell emitting the highest energy photons (in  LS01)  and the Lorentz factor of the outmost shell (in ZP09). 

Additional bursts, and in particular additional bursts containing GeV emission detected by Fermi for which the compactness problem is most efficiently utilized will enable us, hopefully in the near future, to confront the very early afterglow constraint with the lower limits obtained by the compactness problem. Consistency between the two will confirm that we are on the right track towards a resolution of how GRBs work, while a significant contradiction will pose yet another puzzle. 

\section*{Acknowledgments}
We thank R. Sari, and Y. Z. Fan for the helpful discussion.  This
work is supported by an ERC advanced research grant and by the center of excellence in High Energy Astrophysics funded by the Israel Science Foundation by the Schwartzmann chair (TP) and by the National
Natural Science Foundation of China under the grant 10703002 (fYCZ).

\end{document}